\documentclass[11pt,twoside]{article}

\usepackage{asp2006}
\usepackage{epsf}
\usepackage{psfig}
\usepackage{graphicx}
\usepackage{lscape}

\begin{document}

\title{2 types of spicules ``observed" in 3D realistic models}

\author{Juan Mart\'inez-Sykora}

\affil{Institute of Theoretical Astrophysics, University of Oslo, Norway}
\affil{ Lockheed Martin Solar \& Astrophysics Lab, Palo Alto, USA}

\newcommand{\myemail}{juanms@astro.uio.no}
\newcommand{\viscous}{\underline{\underline{\tau}}}
\newcommand{\resistive}{\underline{\underline{\eta}}}
\newcommand{\komment}[1]{\texttt{#1}}

\begin{abstract}

Realistic numerical 3D models of the outer solar atmosphere show two different kind of spicule-like phenomena, as also observed on the solar limb. The numerical models are calculated using the {\em Oslo Staggered Code} (OSC) to solve the full MHD 
equations with non-grey and NLTE radiative transfer and thermal conduction along
the magnetic field lines. The two types of spicules arise as a natural result of the dynamical evolution in the models. We discuss the different properties of these two types of spicules, their differences from observed spicules and what needs to be improved in the models. 

\end{abstract}



\section{Numerical Methods and description of the model}
\label{sec:equations}

The nature of spicules observed at the solar limb have long been a mystery, in this paper we discuss two types of jets that occur naturally in 3D numberical models of the solar atmosphere. The MHD equations are solved in a model spanning the upper convection and corona using the {\it Oslo Stagger Code} (OSC). 
In addition, this code solves a rather realistic NLTE radiative transfer, including scattering, and thermal conduction along the field lines as explained in \citet{paper1}.

The models described below have a grid size of $256\times 128\times 160$
points spanning  $16\times 8\times 16$~Mm$^3$. The grid is uniform in the horizontal direction with a grid spacing of $65$~km. In the vertical direction the grid is non-uniform, 
ensuring that the vertical resolution is good enough to resolve the photosphere and 
transition region with a grid spacing of $32.5$~km, while becoming larger at coronal 
heights. At these resolutions the models have been run for roughly $1.5$ hours solar time. 
We have created two models with different average unsigned field in the photosphere; one with $16$~G (A2) and the other $160$~G (B1). In addition to this ambient field, we introduce a magnetic flux tube into the lower boundary at the bottom boundary \citep{paper1}. 

\section{Results and discussions}
\label{sec:results}

In the models we found two types of spicule-like structures, 
{\it i.e} the so-called type~{\sc i} \citep{Martinez-Sykora:2009kl} and type~{\sc ii} \citep{McIntosh:2007kl}. 

\begin{figure}
\begin{center}
  \includegraphics[width=0.53\textwidth]{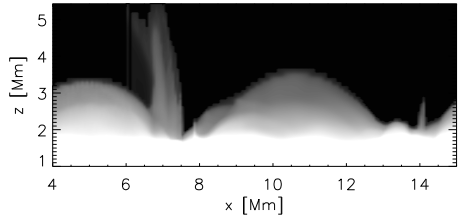}
\end{center}
  \vspace{-0.3cm}
    \caption{\label{fig:calimb} Synthetic image of Ca~{\sc ii} H from the limb of the model. Observe the two types of spicules, type~{\sc i} located at $x=14$~Mm and type~{\sc ii} at $x=7$~Mm. The synthetic image is done with MULTI\_3D.}
      \vspace{-0.3cm}
\end{figure}

A synthetic image of Ca~{\sc ii} at the limb is shown in fig~\ref{fig:calimb} which 
shows the two types of spicules. 
These structures look rather 
similar to what is observed at the limb in the Sun in Ca~{\sc ii}. 

Table~\ref{tab:diff} shows the differences 
between the two types of the spicules in our models and compared to observations. The reader is referred to work that has recently been completed related 
to spicules; \citet{luc2007,De-Pontieu:2007cr,Hansteen+DePontieu2006,Martinez-Sykora:2009kl}.

\begin{table}
\begin{tabular}{ | p{4.2cm} | p{4.2cm} | p{4.2cm}|} 
\hline
{\bf Type~{\sc i}} & {\bf Type~{\sc ii}}  & {\bf Observations}  \\
\hline
\hline
150 examples in both models & 2 examples only in B1 model & Type~{\sc ii} ubiquitous \\ \hline
Length $\approx[0.4,1.5]$~Mm & Length $\approx 5$~Mm & Type~{\sc i} are longer \\ \hline
Duration $\approx[2,5]$~min & Duration $\approx 1$~min & Type~{\sc i} have longer durations \\ \hline
Parabolic profile in time (deceleration) & Complex velocity profiles due to acceleration at different height & Seems to agree (see bibliography) \\ \hline
Up-downflow profile & Only upflow & Seems to agree \\ \hline
Velocities $\approx [5,35]$~km/s & Velocities $\approx$~150~km/s & Type~{\sc i} reach larger velocities\\ \hline
Observed in Ca~{\sc ii} & Counterpart in Transition region emission lines &  Seems to agree \\ \hline
Driven by magneto-acustic shocks & Reconnection & Similar drivers suggested  \\ \hline
\end{tabular}
\caption{\label{tab:diff} Properties of the two types of spicules ``observed" in the models and compared with observations.}
\end{table} 

Most likely, the differences between the two types seem to agree with 
the observations. However, a deeper study needs to be done with the 
type~{\sc ii} spicules found in the model (work in progress). Moreover, a closer comparison with the observations is required. It is interesting to note that the 
appearance of type~{\sc i} does not show a clear preference between models 
with or without flux emergence, while type~{\sc ii} only appear in the model with the largest ambient ambient field (B1), and only after emerging flux
cross the photosphere. Spicules of both types in the models are located at the footpoints of the atmospheric coronal loops, where the field lines are open field lines or at least 
penetrate into the corona. Moreover, the footpoint that is closer to the emerging flux tube is the one that shows most jets. The type~{\sc ii} spicules shows a corresponding 
nearby hot loop which also seems observed in the Sun \citep{De-Pontieu:2009yf}. 
The hot loop ($>10^6$~K) can be observed with coronal emission lines. 

\begin{figure}
\begin{center}
  \includegraphics[width=0.97\textwidth]{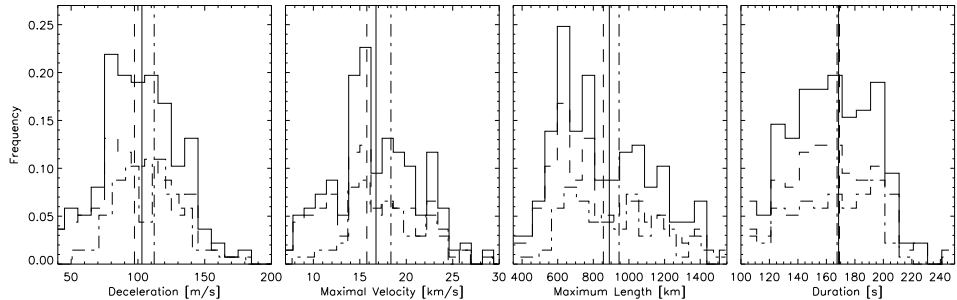}
  \end{center}
    \vspace{-0.3cm}
    \caption{\label{fig:dist} Histograms, normalized to the total number of spicules, for decelerations, maximal velocity, maximum lengths, and duration, from left to right, respectively, measured from the two models (B1 dash and A2 dash-dot line) and the sum (continuum line). The vertical line is the median value of the distribution. The two models show different distributions of the deceleration, maximum lengths, and duration, as well as some differences in maximum velocity. The model B1 shows  on average slightly lower decelerations, shorter length, and shorter velocities than A2.}
      \vspace{-0.3cm}
\end{figure}

In brief, we can summarize the differences between observations and models 
for type~{\sc i} spicules by noting that the upper limits of the deceleration, length, duration, and maximal velocity are smaller in the models \citep{Martinez-Sykora:2009kl}. 
Histograms for deceleration, maximum length, maximal velocity and duration 
for the type~{\sc i} from the model are shown in fig~\ref{fig:dist}. These can 
be compared with the histograms from the observations done by
\citet{De-Pontieu:2007cr}. They show an agreement in the lower part 
to the histograms, and the differences between the B1 and A2 seems similar
to the differences of the two regions observed by \citet{De-Pontieu:2007cr}. 
However, the models do not fit with the upper part of the observed histograms.

In order to improve the models we consider that the resolution of the box is important. The chromosphere is poorly resolved numerically and this affects the size of the structures of the spicules. In addition, low resolution might bring other effects like the diffusion of the shocks (type~{\sc i}) or of the magnetic discontinuity (type~{\sc ii}). With higher resolution we expect sharper shocks, larger range of velocities, better resolved and more frequent spicules.

\begin{figure}
\begin{center}
  \includegraphics[width=0.97\textwidth]{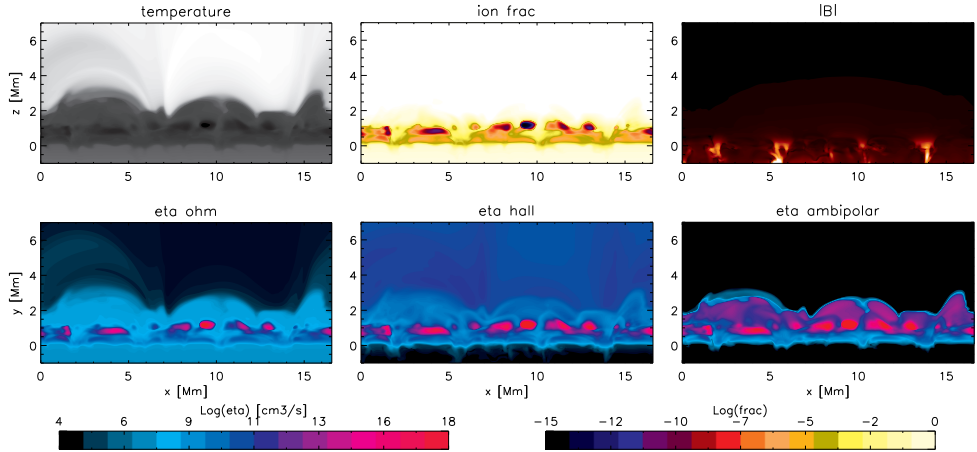}
  \end{center}
  \vspace{-0.3cm}
    \caption{\label{fig:gen} Temperature, ion fraction, magnetic field intensity, and ohm, hall and ambipolar diffusion calculated as post-processing in a 2D cut of the model, from left to right and top to bottom. Observe that Ohm and Hall diffusion are rather important in the lower chromosphere and the ambipolar diffusion in the upper chromosphere.}
      \vspace{-0.3cm}
\end{figure}

In models, it is also important to take time-dependent hydrogen ionization into account in the the upper chromosphere. The ionization of hydrogen in the solar chromosphere and transition region does not obey LTE, or instantaneous statistical equilibrium, as the timescales of ionization and recombination are long compared with HD timescales, especially for magneto-acoustic shocks. The shock temperatures are higher, and the intershock temperatures are lower, in models where time-dependent ionization is considered. This effect will likely change the range of parameter of the spicules type~{\sc i} \citep{Leenaarts:2007sf}. Modeling the chromosphere is strongly important to study properly the radiative losses approximations, NLTE with scattering as has discussed by\citet{Carlsson2010,Leenaarts2010}.

The partial ionization might have other effects on both types of spicules, as well. When considering partial ionization we find that ambipolar diffusion, Hall diffusion and ohmic diffusion contribute at differing rates throughout the chromosphere. The ratio between these three diffusion terms changes from the photosphere up to the transition region (see fig~\ref{fig:gen}).  This will possibly have important effects in the chromosphere as the parameters controlling reconnection and the damping of waves change.

Finally, we note that the range of ambient magnetic field structures that have been modeled only form a small subset of those expected when considering supergranulation, plage and the chromospheric network. In addition, a continuous weak magnetic flux emergence may need to be added, since it has been observed in the models that chromosphere and transition region heights are considerably increased with flux emergence. 

{\bf Acknowledgments:} Inestimable collaboration and contribution with Viggo Hansteen, Bart de Pontieu, Mats Carlsson and Fernando Moreno-Insertis.

\end{document}